\documentclass[aps,prl,reprint,groupedaddress,showkeys]{revtex4-2}

\usepackage{graphicx}%
\usepackage{amsmath,amssymb,amsfonts}%
\usepackage{dcolumn}
\usepackage[title]{appendix}%
\usepackage[utf8]{inputenc}
\usepackage[T1]{fontenc}
\usepackage{textcomp}%
\usepackage{bm}
\usepackage{algpseudocode}%
\usepackage{listings}%

\begin{document}

\title[Observation and control of surface-state formation via optical anisotropy]{Observation and control of potential-dependent surface state formation at a semiconductor--electrolyte interface via the optical anisotropy}

\author{Marco Flieg}
\author{Margot Guidat}
\affiliation{Institute of Physical and Theoretical Chemistry, Universität Tübingen, Auf der Morgenstelle 15, 72076 Tübingen, Germany}

\author{Matthias M. May}
 \email{matthias.may@uni-tuebingen.de}
\affiliation{Institute of Physical and Theoretical Chemistry, Universität Tübingen, Auf der Morgenstelle 15, 72076 Tübingen, Germany}
\affiliation{Center for Light-Matter Interaction, Sensors and Analytics LISA+, Universität Tübingen, Auf der Morgenstelle 15, 72076 Tübingen, Germany}

\date{\today}
\begin{abstract}
{The interface between semiconductors and ion-conducting electrolytes is characterised by charge distributions and potential drops that vary substantially with the evolution of surface states. These surface states at the very interface to the liquid can form or be passivated, depending on the applied potential between electrode and electrolyte, and hereby fundamentally impact properties such as charge transfer. Characterisation and understanding of such potential-dependent surface states with high spatial and temporal resolution is a significant challenge for the understanding and control of semiconductor--electrolyte interfaces. Here, we show that the optical anisotropy of InP(100) can be used to detect the potential-dependent formation of highly ordered surface states under operating conditions. Upon formation of a surface state in the bandgap of the semiconductor, the potential drop and hence the electric field is shifted away from the semiconductor to the Helmholtz-layer of the electrolyte. This modifies the instantaneous response of the optical anisotropy to disturbances of the applied potential. We propose an electrochemical variant of the linear electro-optical effect and our findings open a novel route for understanding these interfaces. The results show how surface states from surface reconstructions at this reactive interface can be switched on or off with the applied potential.}
\end{abstract}


\maketitle

The contact between a semiconductor and an ion-conducting electrolyte, the semiconductor--electrolyte interface, plays an important role in fields such as electrochemical energy conversion, catalysis or sensing applications \cite{Fujishima_photolysis_1972,Kaufman_relevance_band_edges_PEC_2024, Azad_1992}. Here, surface-states at the solid--liquid interface not only impact the electronic properties of the semiconductor, but also the potential distribution in the electrolyte \cite{Gerischer_charge_transfer_sc_electrolyte_1969} and can lead to Fermi-level pinning \cite{Bardeen_fermi-level_pinning_1947}. These surface states are not static and can evolve dynamically over time, with applied potentials, or illumination-induced interaction with the electrolyte \cite{May_coelec_photoelectrosynthetic_interfaces_2022, Favaro_BiVO4_surface_reactions_2018}. However, the access to the (electronic) structure of the solid-liquid interface, the so-called electric double layer (EDL), is not as straightforward as for surfaces in vacuum due to the short mean-free path of electrons, the typical probes in surface science, in an electrolyte. From a materials view, reactive semiconductor--electrolyte interfaces are more complex than metal--electrolyte interfaces, where well-defined single crystal surfaces form the basis for fundamental experiments \cite{Goodwin2024}. This lack of insight into and control of the exact structure of semiconductors with a liquid also impedes the understanding from a perspective of theory \cite{Pham_modelling_interfaces_solar_water_splitting_2017}.

A number of methods exist that deliver complementary information on the quest to understand the electronic and real-space structure of the solid--liquid interface on an atomistic level, but all struggle to deliver information at high temporal and spatial resolution due to the difficulty to convey information through the liquid electrolyte \cite{May_coelec_photoelectrosynthetic_interfaces_2022,Esposito_methods_spatial_photoelectrode_characterisation_2015}. Methods for the time- and potential-dependent observation of surface states are scarce and often involve very elaborate experiments that might lead to deviations from operating -- also called \textit{operando} -- conditions. This is for instance the case for near-ambient pressure X-ray photoelectron spectroscopy due to the very thin electrolyte layer thickness that modifies potential distributions at finite currents. Another method is electrochemical impedance spectroscopy (EIS), but here, the necessary inclusion of a wide frequency range drastically limits the temporal resolution \cite{Hens_photoanodic_dissolution_InP_EIS_2000}. Non-\textit{operando} conditions, where the electrolyte is for instance removed in ultra-high vacuum conditions prior to analysis, will often not tell the full story, as breaking potential control typically leads to rapid structural decay of a given interfacial structure \cite{Loew_InP_RAS_2022}. Here, electrochemical reflection anisotropy spectroscopy (EC-RAS) is an emerging, complementary method in this toolkit with extremely high (sub-nm) interface sensitivity and a temporal resolution in the order of ms \cite{Aspnes1985,Guidat_EC-RAS_review_2023}. The anisotropy of the dielectric tensor of a surface can render RAS sensitive to interfacial electric fields via the linear electro-optic effect \cite{Leo_PRB_1989}, but it was so far unclear, if this can be exploited for the study of solid--liquid interfaces.

\begin{figure}[h!]
    \centering
    \includegraphics[width=0.5\textwidth]{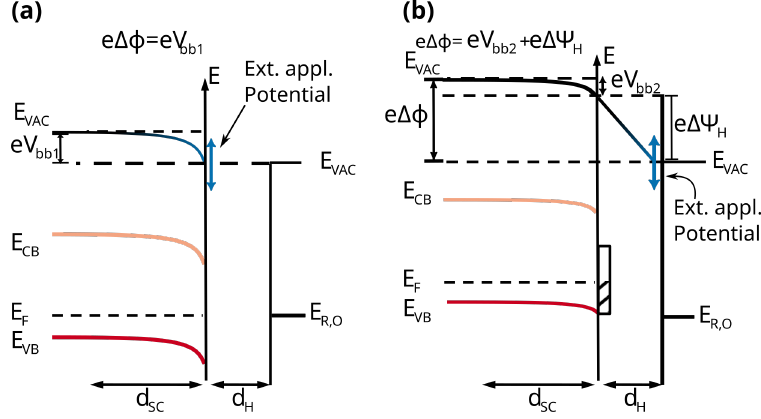}
    \caption{Band diagrams of a p-type semiconductor in contact with an electrolyte. (a) Semiconductor in contact with an electrolyte and no surface states. (b) Same situation as (a), but with a high density of surface states (vertical box, striped for occupied), shifting the potential drop from the solid to the Helmholtz-layer, $e\Delta\Psi_H$. A variation of the applied potential leads to a system response (blue arrows) in (a) by a change in band-bending and (b) a change in the Helmholtz-layer potential drop.}
    \label{fig1:Band_diagramm}
\end{figure}

Figure \ref{fig1:Band_diagramm} shows the simplified -- neglecting dipole layers -- band diagrams of a p-doped semiconductor in contact with an electrolyte without (a) and with (b) active surface states. In the case without surface states, the equilibration of electrochemical potentials, i.e. the redox-level, $E_{R,O}$, in the electrolyte and the Fermi-level, $E_F$, in the solid together with the fulfilment of the charge-neutrality condition lead to a depletion layer in the semiconductor. This means that the potential drop, i.e. the difference in work function between semiconductor and redox-couple, which is indicated by the band-bending, $e\Delta\Phi=eV_{bb}$, occurs exclusively in the semiconductor. This situation is changed by the introduction of surface states, which are occupied by additional charge, $Q_{ss}$, leading to Fermi-level pinning at the semiconductor. Now part of the potential drop occurs within the Helmholtz-layer and potentials applied to the electrode will -- in a certain potential range -- only change this potential drop. This means that the electric fields over the solid-liquid interface and their evolution with applied potentials depend on the density of active surface states \cite{Jaegermann_modern_asp_elchem_1996, May_coelec_photoelectrosynthetic_interfaces_2022}. A time- and potential-resolved access to the electric field distribution as a function of applied potentials would therefore allow to understand the dynamic formation of surface states in electrochemical environments.

In this work, we explore the interface between InP(100) and an aqueous electrolyte by electrochemical reflection anisotropy spectroscopy in a potential- and time-resolved manner. We find that the response of the optical anisotropy to varied applied potentials directly allows to follow the formation of surface states at the very interface via the linear electro-optic effect. Depending on the applied potential, highly ordered surfaces with active or inactive electronic states within or outside of the bandgap region can be switched on or off. This lays the ground for a novel approach to study and, due to the high temporal resolution of the optical probe, control surface-state evolution at semiconductor--liquid interfaces.

The starting point of the experiments is indium phosphide (InP), covered by an ``epi-ready'' oxide. In dilute (10\,mM) hydrochloric acid (HCl), this oxide layer can be dissolved under cathodic potentials and a well-ordered interface develops \cite{Loew_InP_RAS_2022}. As we illuminate the sample with the white light from our spectrometer, a splitting of the Quasi-Fermi levels of holes and electrons develops, giving rise to a photovoltage-induced shift of band positions. Here and in the following, we assume for the sake of simplicity this photo-induced shift to be constant. Under this assumption, we use one Fermi-level, $E_F$, for the discussion and interpretation of the results. In acidic aqueous conditions, the $E_F$ of the semiconductor aligns with the level for proton reduction (hydrogen evolution). This energetic alignment upon contact of the two phases is achieved by a charge re-distribution in both, the semiconductor and the electrolyte, leading to band-bending and an electric field within the semiconductor. If an external potential is applied to the semiconductor electrode, band bending and hereby the interfacial electric field will change as a function of the applied potential (blue arrow in Fig.~\ref{fig1:Band_diagramm}a). This situation is drastically changed by the introduction of surface states with energy levels within the forbidden gap (Fig.~\ref{fig1:Band_diagramm}b), which can arise from initial corrosion or the formation of adsorbates on the interface. Similar to semiconductor--metal interfaces \cite{Bardeen_fermi-level_pinning_1947,Tersoff1985}, the Fermi-level is now pinned in a certain potential range, the band bending remains fixed and the potential drop is shifted towards the electrolyte side into the Helmholtz double-layer \cite{Jaegermann_modern_asp_elchem_1996,May_coelec_photoelectrosynthetic_interfaces_2022}. This also means that the electric field in the semiconductor remains constant upon a change of the applied potential.

\begin{figure*}[ht]
    \centering
    \includegraphics[width=\linewidth]{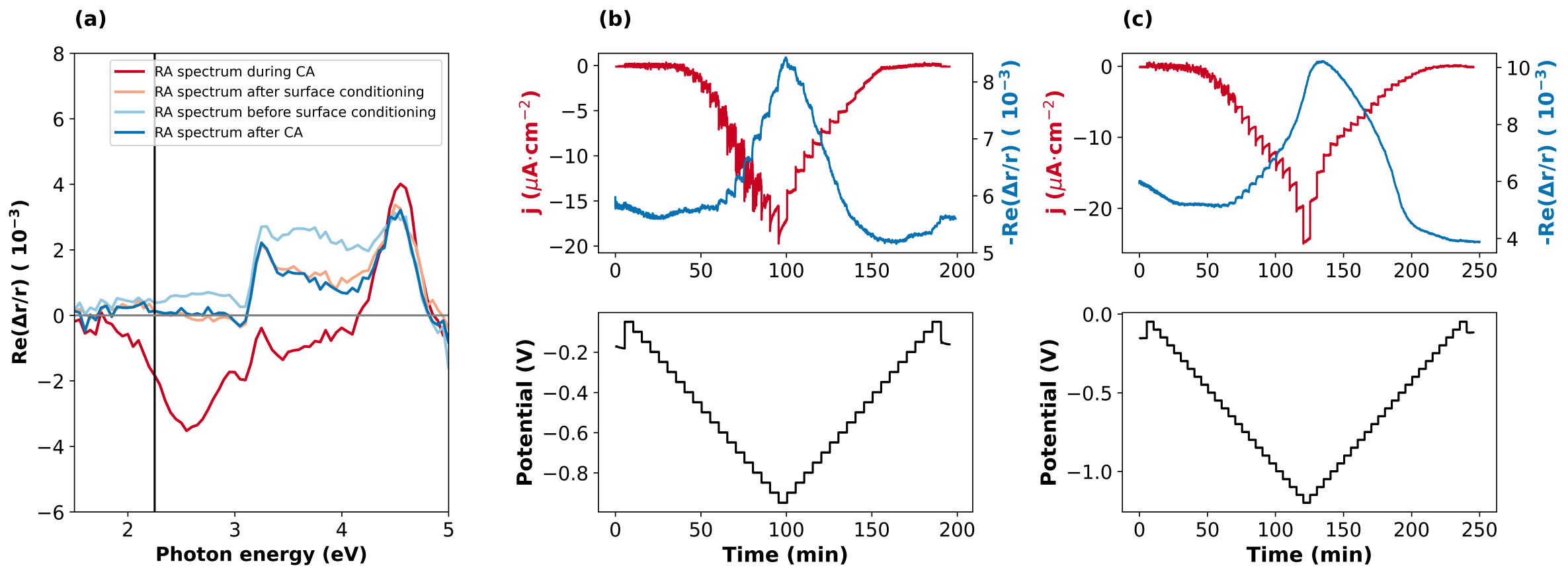}
    \caption{Optical response of the system as a function of the applied potential. (a) Reflection anisotropy spectra in the electrolyte during and after chronoamperometry (CA) at -0.9\,V vs.~the Ag/AgCl reference electrode (RE) as well as before and after conditioning by potential cycling. The vertical line marks the photon energy selected for the transients. 
    (b) Response of the system in the potential range without surface states. Upper panel: Current density (red) and RA-transient (blue) at 2.25\,eV photon energy following the applied potential steps (lower panel) of 50\,mV/5\,min between OCP and -0.95\,V vs.~RE. (c) Response of the system in the extended potential range.}
    \label{fig2:spectra_transients}
\end{figure*}

Our spectroelectrochemical setup probes this situation with optical reflection anisotropy spectroscopy (RAS) in a near-normal reflection geometry through the electrolyte, while the potential of the InP(100) sample with respect to the electrolyte is controlled in a three-electrode configuration. As the bulk of InP(100) is optically isotropic, the signal arises solely from the optical anisotropy, $\Delta r/r$, of the very solid--liquid interface and is therefore only non-zero if the interface exhibits a minimum degree of ordering \cite{Guidat_EC-RAS_review_2023}. For InP(100) in contact with aqueous electrolytes, an electrochemical configuration space exists in a narrow region with respect to potential and electrolyte concentration, where such an ordering occurs, which is in turn associated with a characteristic RA spectrum \cite{Loew_InP_RAS_2022}. The parallel acquisition of RA spectra during a linear potential sweep then allows to identify spectral signatures of potential-dependent surface phases (Fig.~\ref{fig2:spectra_transients}a). For improved temporal resolution in transient RA spectroscopy, a photon energy of interest is fixed and the anisotropy signal is recorded over time (Fig.~\ref{fig2:spectra_transients}b,c). This mode does, for instance, allow to probe surface thermodynamics and the impact of defects thereon \cite{May_time-resolved_water_adsorption_2019, Vazquez-Miranda_adsorbate_isotherm_RAS_Cu110_HCl_2020} or the structural evolution of short-lived interfacial structures upon the application of potential steps.

\begin{figure}[ht]
    \centering
    \includegraphics[width=\linewidth]{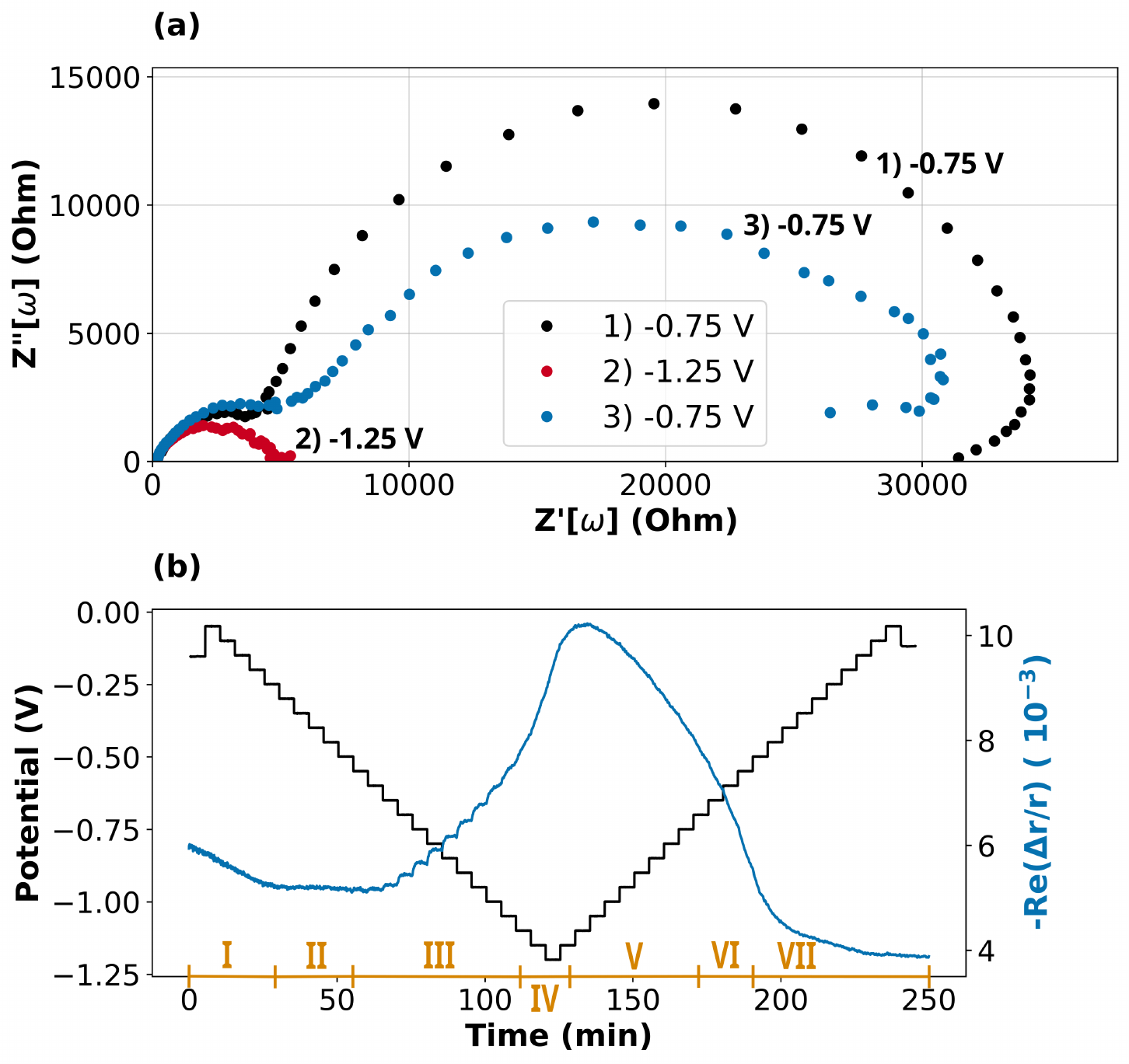}
    \caption{Evolution of surface states and optical anisotropy in the different potential regimes. (a) Nyquist plots from the three electrochemical impedance measurements (EIS) conducted in one experiment sequence in which the following potential steps of 18\,min each: first -0.75\,V, second -1.25\,V, and finally back to -0.75\,V. The changed half-cycles show the increased density of surface states from a semiconducting behaviour (1) to metal-like properties (2), and -- in a partially reversible manner -- back to semiconducting characteristics (3). (b) The RA transient at 2.25\,eV (blue) is plotted together with the applied potential (black). The aforementioned potential regimes are highlighted in orange: (I) the RA is decreasing, (II) it is constant, (III) rising with fast response, (IV) the RA is increasing without steps due to the constant band bending, (V) the RAS is decreasing without steps, (VI) the RA is decreasing with subdued steps, and (VII) finally without steps.}
    \label{fig3:Nyquist_and_Regimes}
\end{figure}

We now combine potential steps and sweeps by scanning a given potential range with a series of potential steps, allowing to discriminate between fast and slow processes at the interface in the RA transient (Fig.~\ref{fig2:spectra_transients}b,c). In the first case, we limit the potential range to cathodic potentials where no significant formation of surface states in the bandgap is to be expected \cite{Euchner_phase_diagram_InP_2025}. The RA transient follows these steps with two different time-constants, fast and slow processes impacting the optical anisotropy. Especially in the range -0.5 to -0.7\,V (vs.~Ag/AgCl), there is an instant increase of the anisotropy, followed by a much slower decrease. Beyond -0.7\,V, however, the slower process also contributes to a positive increase of the anisotropy. Reversing the direction of the potential scan, these effects still persist, but not as pronounced, anymore. In Fig.~\ref{fig2:spectra_transients}c, we extend the cathodic potential range and observe that during the forward potential scan beyond -0.95\,V, the RA transient does not show a stepwise behaviour, anymore, but continuously increases. During the backward scan, there remains only a barely resolved discontinuity of the RA transient at the very potential steps. Electrochemical impedance spectroscopy (Fig.~\ref{fig3:Nyquist_and_Regimes}a) shows that these different potential regimes are associated with distinctly different interfacial capacitances and surface states \cite{Ponomarev_EIS_InP_1995,Hens_photoanodic_dissolution_InP_EIS_2000}. While the initial surface at -0.75\,V shows two very distinct half-circles, only one half-circle remains at the cathodic extremum of -1.25\,V. The initial shape of the impedance spectrum is qualitatively restored by going back to -0.75\,V, but the second half-circle is subdued, indicating an only partial reversibility.

This observation shows that a fast response of the optical anisotropy -- i.e. steps in RA transients concurrent with potential steps -- can be observed in a potential range up to -0.95\,V. This fast response does, however, disappear for cathodic potentials beyond this threshold, when an in-gap surface state, that also modifies the interfacial capacitance, develops. The observed potential range aligns with results from recent computational work assessing the most stable InP(100)-derived surface structures containing H and Cl by means of the computational hydrogen electrode \cite{Euchner_phase_diagram_InP_2025}: At moderately cathodic potentials vs.~Ag/AgCl, the $(2\times2)$-6H structure is expected to be stable, which does not exhibit electronic states within the bandgap. At more cathodic potentials, however, structures such as the $(2\times2)$-mixed-4H structure start to dominate, where surface states within the bandgap exist.

The slower reactions of the optical anisotropy to the externally applied potentials can be understood in the view of surface restructuring and ion ad-/desorption. In the first regime I, from OCP to -0.25\,V, the anisotropy changes linearly with the applied potential, which can be either due to  very slow surface re-structuring or ion ad-/desoprtion. Indeed, in this region, a surface state that reduces the band gap, but does not make the surface metallic, is expected \cite{Euchner_phase_diagram_InP_2025}. In region II, the anisotropy stays nearly constant, indicating a stable surface structure. Beyond -0.55\,V, however, the overall anisotropy shows an increase with a fast and a slow contribution, which indicates that, in addition to the effect of the electric field, a more slow process such as ion adsorption takes place, which would be expected for an increasing potential drop in the Helmholtz-layer. In region IV, however, the slope of the RA signal change with potential becomes steeper, but no RA steps can be observed, anymore. This is an indicator of a continued surface restructuring and the expected $(2\times2)$-mixed-4H surface would indeed be metallic \cite{Euchner_phase_diagram_InP_2025}. When the potential scan is reversed, both the RA signal and the current are reversing more sluggishly, which indicates a hysteresis effect of the surface restructuring, combined with a partial non-reversibility as indicated by the EIS data. 

The relation between fast response of the optical anisotropy and the absence of in-gap surface states can be understood with the linear electro-optic effect (LEO) \cite{Kyser1970,Leo_PRB_1989}. In general, the linear, field-induced changes on the components $\epsilon_{ij}$ of the dielectric function of a semiconductor are given as

\begin{equation}
 \Delta\epsilon_{ij} = Z_{ijk}F_k
 \end{equation}

where $\mathbf{F}$ is the electric field and $Z_{ijk}$ are the components of the tensor describing the linear electro-optic effect \cite{Leo_PRB_1989}. The field-induced optical anisotropy, $\Delta r_F/r$, for normal incidence on the InP(100) surface with light polarization direction, $\theta$, measured counter-clockwise from the [001] direction then becomes

 \begin{equation}
 \Delta r_F/r = \sin(2\theta)\times\left[g_0(E)|\mathbf{F}(V_{bb})| + g_1(E)|\mathbf{F}(V_{appl})|\right]
\end{equation}

where $g(E)$ is a function of the crystal structure and the probing photon energy. In the case of semiconductor--electrolyte interface, we have to consider that $\mathbf{F}(V_{bb},V_{appl})$ is a function of initial band bending, $V_{bb}$, and the externally applied potential, $V_{appl}$, which in turn can also modify the potential drop in the Helmholtz-layer, $\Delta\Psi_H$ (Fig.~\ref{fig1:Band_diagramm}). We therefore introduce here the distinction between $g_0(E)$ and $g_1(E)$, as the band bending-induced field, $\mathbf{F}(V_{bb})$, will mostly impact the solid, surface and near-surface region of the semiconductor as described by $g_0$ in the classical linear electro-optic effect \cite{Leo_PRB_1989}. The field introduced by the applied potential, $\mathbf{F}(V_{appl})$, on the other hand, will either modify $V_{bb}$, which means that $g_1=g_0$ applies, or $\Delta\Psi_H$, which means that only the very solid surface and possibly adsorbates in the Helmholtz-layer as described by $g_1\neq g_0$. While $g_1$ might be small in magnitude, the EDL-LEO can be still significant due to the large electric fields in the order of up to MV/m that arise at the EDL \cite{Baldelli2005}. 

To summarise, our work shows that highly ordered interfaces between a semiconductor and an electrolyte can be prepared as a function of the applied potential and the associated electronic surface states can be hereby switched on and off. The formation or passivation of active surface states within the bandgap can be probed by the electrochemical variant of the linear electro-optic effect on the optical anisotropy of semiconductor--electrolyte interfaces that we propose here. This opens a novel approach to understand surface state formation and electric field effects at the electric double layer. The highly ordered nature of the interfaces will also make them accessible to electronic structure modelling approaches that are based on small supercells. These findings could also help understanding and controlling other semiconductor--electrolyte interfaces, where for instance prior to co-catalyst electrodeposition, surface states are typically to be avoided in order to prevent surface charge-carrier recombination. Unlike other optical methods such as sum frequency generation that can also be used to study surface states \cite{Baldelli2005}, RAS is a linear optical spectroscopy, requiring only moderate light intensities and working in a normal incidence reflection geometry, which also allows to study systems prone to radiation damage. The approach could therefore for instance also be used to control the solution-based growth of perovskite-based solar absorbers \cite{Yang_surface_recombination_solution-grown_perovskite_single-crystal_2015}, already assessing and minimising the formation of surface states during growth. The next steps for more in-depth studies would be the modelling of the computational spectra and probing the exact structures experimentally for instance with electrochemical scanning tunnelling microscopy.

\section{Acknowledgments}
\begin{acknowledgments}
This work was funded by the German Research Foundation (DFG) under project number 434023472 and the German Bundesministerium für Bildung and Forschung (BMBF), project ``H2Demo'' (No.~03SF0619K). We thank H.~Euchner for discussions, M.~Diecke and E.A.~Schmitt for assistance with sample preparation.
\end{acknowledgments}

\subsection*{Data availability}

 The raw data used for this study will be made openly accessible on Zenodo \cite{Flieg_dataset_2025} upon acceptance of the manuscript.

\subsection*{Author contributions}

M.F. carried out the experiments and analysed the data. M.M.M., M.G., and M.F. designed the experiments, M.M.M. and M.G. supervised the work. M.F. and M.M.M. wrote the manuscript. All authors discussed the data and commented on the manuscript.

\appendix
\section{Appendix: Methods}

\subsection*{Spectroelectrochemical setup}

The experimental setup consists of a Princeton Applied Research VersaSTAT 3F potentiostat, a Laytec EpiRAS spectrometer, and a diffused LED light ring. All components are added and mounted to the Faraday cage. The experiments were conducted in a photoelectrochemical Schlenk cell \cite{Schmitt_PEC_Schlenk_cell_2023} with taper joints as inlets and outlets, and a quartz glass window for optical access with the spectrometer. By working under Schlenk conditions, an oxygen-, dust-, and particle-free environment can be assured. 

All measurements were conducted in a three-electrode setup. The working electrodes (WE) were the InP samples, a Pt-wire the counter electrode (CE), and Ag/AgCl in 3.5\,M KCl was used as reference electrode (RE). 

Pre-conditioning of the samples to remove the initial oxide layer was performed with cyclic voltammetry in a potential range from -1 to -0.5\,V with a slew rate of 20\,mV/s for 10 cycles. The initial and the end potentials were set to 0\,V vs.~OCP and the potentials from the potential range were set vs.~the reference electrode. For EIS, the measurements were conducted from 100\,kHz to 10\,mHz. 

For the electrolyte, a 32\% HCl solution (CAS-No. 7647-01-0) was diluted with Milli-Q grade water. This water was also used for all cleaning steps and to set up the KCl solutions from 99.997\% KCl (CAS-No. 7447-40-7, purchased from Alfa Aesar). All chemicals were used without further purification. The samples used in this study were $\sim$1x1\,cm$^2$ pieces of a p-doped InP(100) wafer with a 2$^\circ$ offcut from InPACT. The polished topside was covered by a chemically produced ``epi-ready'' oxide layer.

\subsection*{Sample preparation}
Samples, where no EIS measurements were performed, were prepared as follows: In a 16\,cm glass tube a wire of 19\,cm was led through of which both ends were stripped of the insulation. The oxide layer on the backside of the InP wafer pieces (1\,cm$^2$ squared) was mechanically removed with a scalpel for contacting. The wire was connected to the backside with a drop of silver conductive resin, followed by curing for 15\,minutes. Then, the junction of the wafer and glass rod, the sample backside, and the sample edges were covered with an isolating, 2-component epoxy resin and cured for 12\,h. This procedure was repeated one more time. 

For the samples, where EIS was to be performed, a sleeve of PVDF was 3D-printed (Apium P-220) and the wafer piece placed inside the sleeve. By heating the protruding edges, the sample was melted into the PVDF sleeve. A wire was placed upon the scratched backside with capton tape and a PEEK rod screwed into the backside of the PVDF sleeve.


\begin{thebibliography}{26}%
\makeatletter
\providecommand \@ifxundefined [1]{%
 \@ifx{#1\undefined}
}%
\providecommand \@ifnum [1]{%
 \ifnum #1\expandafter \@firstoftwo
 \else \expandafter \@secondoftwo
 \fi
}%
\providecommand \@ifx [1]{%
 \ifx #1\expandafter \@firstoftwo
 \else \expandafter \@secondoftwo
 \fi
}%
\providecommand \natexlab [1]{#1}%
\providecommand \enquote  [1]{``#1''}%
\providecommand \bibnamefont  [1]{#1}%
\providecommand \bibfnamefont [1]{#1}%
\providecommand \citenamefont [1]{#1}%
\providecommand \href@noop [0]{\@secondoftwo}%
\providecommand \href [0]{\begingroup \@sanitize@url \@href}%
\providecommand \@href[1]{\@@startlink{#1}\@@href}%
\providecommand \@@href[1]{\endgroup#1\@@endlink}%
\providecommand \@sanitize@url [0]{\catcode `\\12\catcode `\$12\catcode
  `\&12\catcode `\#12\catcode `\^12\catcode `\_12\catcode `\%12\relax}%
\providecommand \@@startlink[1]{}%
\providecommand \@@endlink[0]{}%
\providecommand \url  [0]{\begingroup\@sanitize@url \@url }%
\providecommand \@url [1]{\endgroup\@href {#1}{\urlprefix }}%
\providecommand \urlprefix  [0]{URL }%
\providecommand \Eprint [0]{\href }%
\providecommand \doibase [0]{https://doi.org/}%
\providecommand \selectlanguage [0]{\@gobble}%
\providecommand \bibinfo  [0]{\@secondoftwo}%
\providecommand \bibfield  [0]{\@secondoftwo}%
\providecommand \translation [1]{[#1]}%
\providecommand \BibitemOpen [0]{}%
\providecommand \bibitemStop [0]{}%
\providecommand \bibitemNoStop [0]{.\EOS\space}%
\providecommand \EOS [0]{\spacefactor3000\relax}%
\providecommand \BibitemShut  [1]{\csname bibitem#1\endcsname}%
\let\auto@bib@innerbib\@empty
\bibitem [{\citenamefont {Fujishima}\ and\ \citenamefont
  {Honda}(1972)}]{Fujishima_photolysis_1972}%
  \BibitemOpen
  \bibfield  {author} {\bibinfo {author} {\bibfnamefont {A.}~\bibnamefont
  {Fujishima}}\ and\ \bibinfo {author} {\bibfnamefont {K.}~\bibnamefont
  {Honda}},\ }\bibfield  {title} {\bibinfo {title} {{Electrochemical Photolysis
  of Water at a Semiconductor Electrode}},\ }\href
  {https://doi.org/10.1038/238037a0} {\bibfield  {journal} {\bibinfo  {journal}
  {Nature}\ }\textbf {\bibinfo {volume} {238}},\ \bibinfo {pages} {37}
  (\bibinfo {year} {1972})}\BibitemShut {NoStop}%
\bibitem [{\citenamefont {Kaufman}\ \emph {et~al.}()\citenamefont {Kaufman},
  \citenamefont {Nielander}, \citenamefont {Meyer}, \citenamefont {Maldonado},
  \citenamefont {Ardo},\ and\ \citenamefont
  {Boettcher}}]{Kaufman_relevance_band_edges_PEC_2024}%
  \BibitemOpen
  \bibfield  {author} {\bibinfo {author} {\bibfnamefont {A.~J.}\ \bibnamefont
  {Kaufman}}, \bibinfo {author} {\bibfnamefont {A.~C.}\ \bibnamefont
  {Nielander}}, \bibinfo {author} {\bibfnamefont {G.~J.}\ \bibnamefont
  {Meyer}}, \bibinfo {author} {\bibfnamefont {S.}~\bibnamefont {Maldonado}},
  \bibinfo {author} {\bibfnamefont {S.}~\bibnamefont {Ardo}},\ and\ \bibinfo
  {author} {\bibfnamefont {S.~W.}\ \bibnamefont {Boettcher}},\ }\bibfield
  {title} {\bibinfo {title} {{Absolute band-edge energies are over-emphasized
  in the design of photoelectrochemical materials}},\ }\href
  {https://doi.org/10.1038/s41929-024-01161-0} {\bibfield  {journal} {\bibinfo
  {journal} {Nature Catalysis}\ }\textbf {\bibinfo {volume} {7}},\ \bibinfo
  {pages} {615}}\BibitemShut {NoStop}%
\bibitem [{\citenamefont {Azad}\ \emph {et~al.}(1992)\citenamefont {Azad},
  \citenamefont {Akbar}, \citenamefont {Mhaisalkar}, \citenamefont
  {Birkefeld},\ and\ \citenamefont {Goto}}]{Azad_1992}%
  \BibitemOpen
  \bibfield  {author} {\bibinfo {author} {\bibfnamefont {A.~M.}\ \bibnamefont
  {Azad}}, \bibinfo {author} {\bibfnamefont {S.~A.}\ \bibnamefont {Akbar}},
  \bibinfo {author} {\bibfnamefont {S.~G.}\ \bibnamefont {Mhaisalkar}},
  \bibinfo {author} {\bibfnamefont {L.~D.}\ \bibnamefont {Birkefeld}},\ and\
  \bibinfo {author} {\bibfnamefont {K.~S.}\ \bibnamefont {Goto}},\ }\bibfield
  {title} {\bibinfo {title} {{Solid‐State Gas Sensors: A Review}},\ }\href
  {https://doi.org/10.1149/1.2069145} {\bibfield  {journal} {\bibinfo
  {journal} {Journal of The Electrochemical Society}\ }\textbf {\bibinfo
  {volume} {139}},\ \bibinfo {pages} {3690} (\bibinfo {year}
  {1992})}\BibitemShut {NoStop}%
\bibitem [{\citenamefont
  {Gerischer}(1969)}]{Gerischer_charge_transfer_sc_electrolyte_1969}%
  \BibitemOpen
  \bibfield  {author} {\bibinfo {author} {\bibfnamefont {H.}~\bibnamefont
  {Gerischer}},\ }\bibfield  {title} {\bibinfo {title} {{Charge transfer
  processes at semiconductor-electrolyte interfaces in connection with problems
  of catalysis}},\ }\href {https://doi.org/16/0039-6028(69)90269-6} {\bibfield
  {journal} {\bibinfo  {journal} {Surface Science}\ }\textbf {\bibinfo {volume}
  {18}},\ \bibinfo {pages} {97} (\bibinfo {year} {1969})}\BibitemShut {NoStop}%
\bibitem [{\citenamefont {Bardeen}()}]{Bardeen_fermi-level_pinning_1947}%
  \BibitemOpen
  \bibfield  {author} {\bibinfo {author} {\bibfnamefont {J.}~\bibnamefont
  {Bardeen}},\ }\bibfield  {title} {\bibinfo {title} {Surface states and
  rectification at a metal semi-conductor contact},\ }\href
  {https://doi.org/10.1103/physrev.71.717} {\bibfield  {journal} {\bibinfo
  {journal} {Physical Review}\ }\textbf {\bibinfo {volume} {71}},\ \bibinfo
  {pages} {717}}\BibitemShut {NoStop}%
\bibitem [{\citenamefont {May}\ and\ \citenamefont
  {Jaegermann}(2022)}]{May_coelec_photoelectrosynthetic_interfaces_2022}%
  \BibitemOpen
  \bibfield  {author} {\bibinfo {author} {\bibfnamefont {M.~M.}\ \bibnamefont
  {May}}\ and\ \bibinfo {author} {\bibfnamefont {W.}~\bibnamefont
  {Jaegermann}},\ }\bibfield  {title} {\bibinfo {title} {Combining experimental
  and computational methods to unravel the dynamical structure of
  photoelectrosynthetic interfaces},\ }\href
  {https://doi.org/10.1016/j.coelec.2022.100968} {\bibfield  {journal}
  {\bibinfo  {journal} {Current Opinion in Electrochemistry}\ }\textbf
  {\bibinfo {volume} {34}},\ \bibinfo {pages} {100968} (\bibinfo {year}
  {2022})}\BibitemShut {NoStop}%
\bibitem [{\citenamefont {Favaro}\ \emph {et~al.}(2018)\citenamefont {Favaro},
  \citenamefont {Abdi}, \citenamefont {Lamers}, \citenamefont {Crumlin},
  \citenamefont {Liu}, \citenamefont {van~de Krol},\ and\ \citenamefont
  {Starr}}]{Favaro_BiVO4_surface_reactions_2018}%
  \BibitemOpen
  \bibfield  {author} {\bibinfo {author} {\bibfnamefont {M.}~\bibnamefont
  {Favaro}}, \bibinfo {author} {\bibfnamefont {F.~F.}\ \bibnamefont {Abdi}},
  \bibinfo {author} {\bibfnamefont {M.}~\bibnamefont {Lamers}}, \bibinfo
  {author} {\bibfnamefont {E.~J.}\ \bibnamefont {Crumlin}}, \bibinfo {author}
  {\bibfnamefont {Z.}~\bibnamefont {Liu}}, \bibinfo {author} {\bibfnamefont
  {R.}~\bibnamefont {van~de Krol}},\ and\ \bibinfo {author} {\bibfnamefont
  {D.~E.}\ \bibnamefont {Starr}},\ }\bibfield  {title} {\bibinfo {title}
  {{Light-Induced Surface Reactions at the Bismuth Vanadate/Potassium Phosphate
  Interface}},\ }\href {https://doi.org/10.1021/acs.jpcb.7b06942} {\bibfield
  {journal} {\bibinfo  {journal} {Journal of Physical Chemistry B}\ }\textbf
  {\bibinfo {volume} {122}},\ \bibinfo {pages} {801} (\bibinfo {year}
  {2018})}\BibitemShut {NoStop}%
\bibitem [{\citenamefont {Goodwin}\ \emph {et~al.}(2024)\citenamefont
  {Goodwin}, \citenamefont {L{\"o}mker}, \citenamefont {Degerman},
  \citenamefont {Davies}, \citenamefont {Shipilin}, \citenamefont
  {Garcia-Martinez}, \citenamefont {Koroidov}, \citenamefont {Katja~Mathiesen},
  \citenamefont {Rameshan}, \citenamefont {Rodrigues}, \citenamefont
  {Schlueter}, \citenamefont {Amann},\ and\ \citenamefont
  {Nilsson}}]{Goodwin2024}%
  \BibitemOpen
  \bibfield  {author} {\bibinfo {author} {\bibfnamefont {C.~M.}\ \bibnamefont
  {Goodwin}}, \bibinfo {author} {\bibfnamefont {P.}~\bibnamefont {L{\"o}mker}},
  \bibinfo {author} {\bibfnamefont {D.}~\bibnamefont {Degerman}}, \bibinfo
  {author} {\bibfnamefont {B.}~\bibnamefont {Davies}}, \bibinfo {author}
  {\bibfnamefont {M.}~\bibnamefont {Shipilin}}, \bibinfo {author}
  {\bibfnamefont {F.}~\bibnamefont {Garcia-Martinez}}, \bibinfo {author}
  {\bibfnamefont {S.}~\bibnamefont {Koroidov}}, \bibinfo {author}
  {\bibfnamefont {J.}~\bibnamefont {Katja~Mathiesen}}, \bibinfo {author}
  {\bibfnamefont {R.}~\bibnamefont {Rameshan}}, \bibinfo {author}
  {\bibfnamefont {G.~L.~S.}\ \bibnamefont {Rodrigues}}, \bibinfo {author}
  {\bibfnamefont {C.}~\bibnamefont {Schlueter}}, \bibinfo {author}
  {\bibfnamefont {P.}~\bibnamefont {Amann}},\ and\ \bibinfo {author}
  {\bibfnamefont {A.}~\bibnamefont {Nilsson}},\ }\bibfield  {title} {\bibinfo
  {title} {{Operando probing of the surface chemistry during the Haber–Bosch
  process}},\ }\href {https://doi.org/10.1038/s41586-023-06844-5} {\bibfield
  {journal} {\bibinfo  {journal} {Nature}\ }\textbf {\bibinfo {volume} {625}},\
  \bibinfo {pages} {282} (\bibinfo {year} {2024})}\BibitemShut {NoStop}%
\bibitem [{\citenamefont {Pham}\ \emph {et~al.}(2017)\citenamefont {Pham},
  \citenamefont {Ping},\ and\ \citenamefont
  {Galli}}]{Pham_modelling_interfaces_solar_water_splitting_2017}%
  \BibitemOpen
  \bibfield  {author} {\bibinfo {author} {\bibfnamefont {T.~A.}\ \bibnamefont
  {Pham}}, \bibinfo {author} {\bibfnamefont {Y.}~\bibnamefont {Ping}},\ and\
  \bibinfo {author} {\bibfnamefont {G.}~\bibnamefont {Galli}},\ }\bibfield
  {title} {\bibinfo {title} {Modelling heterogeneous interfaces for solar water
  splitting},\ }\href {https://doi.org/10.1038/nmat4803} {\bibfield  {journal}
  {\bibinfo  {journal} {Nature Materials}\ }\textbf {\bibinfo {volume} {16}},\
  \bibinfo {pages} {401} (\bibinfo {year} {2017})}\BibitemShut {NoStop}%
\bibitem [{\citenamefont {Esposito}\ \emph {et~al.}(2015)\citenamefont
  {Esposito}, \citenamefont {Baxter}, \citenamefont {John}, \citenamefont
  {Lewis}, \citenamefont {Moffat}, \citenamefont {Ogitsu}, \citenamefont
  {O{'}Neil}, \citenamefont {Pham}, \citenamefont {Talin}, \citenamefont
  {Velazquez},\ and\ \citenamefont
  {Wood}}]{Esposito_methods_spatial_photoelectrode_characterisation_2015}%
  \BibitemOpen
  \bibfield  {author} {\bibinfo {author} {\bibfnamefont {D.~V.}\ \bibnamefont
  {Esposito}}, \bibinfo {author} {\bibfnamefont {J.~B.}\ \bibnamefont
  {Baxter}}, \bibinfo {author} {\bibfnamefont {J.}~\bibnamefont {John}},
  \bibinfo {author} {\bibfnamefont {N.~S.}\ \bibnamefont {Lewis}}, \bibinfo
  {author} {\bibfnamefont {T.~P.}\ \bibnamefont {Moffat}}, \bibinfo {author}
  {\bibfnamefont {T.}~\bibnamefont {Ogitsu}}, \bibinfo {author} {\bibfnamefont
  {G.~D.}\ \bibnamefont {O{'}Neil}}, \bibinfo {author} {\bibfnamefont {T.~A.}\
  \bibnamefont {Pham}}, \bibinfo {author} {\bibfnamefont {A.~A.}\ \bibnamefont
  {Talin}}, \bibinfo {author} {\bibfnamefont {J.~M.}\ \bibnamefont
  {Velazquez}},\ and\ \bibinfo {author} {\bibfnamefont {B.~C.}\ \bibnamefont
  {Wood}},\ }\bibfield  {title} {\bibinfo {title} {Methods of photoelectrode
  characterization with high spatial and temporal resolution},\ }\href
  {https://doi.org/10.1039/C5EE00835B} {\bibfield  {journal} {\bibinfo
  {journal} {Energy \& Environmental Science}\ }\textbf {\bibinfo {volume}
  {8}},\ \bibinfo {pages} {2863} (\bibinfo {year} {2015})}\BibitemShut
  {NoStop}%
\bibitem [{\citenamefont {Hens}\ and\ \citenamefont
  {Gomes}(2000)}]{Hens_photoanodic_dissolution_InP_EIS_2000}%
  \BibitemOpen
  \bibfield  {author} {\bibinfo {author} {\bibfnamefont {Z.}~\bibnamefont
  {Hens}}\ and\ \bibinfo {author} {\bibfnamefont {W.~P.}\ \bibnamefont
  {Gomes}},\ }\bibfield  {title} {\bibinfo {title} {Photoanodic dissolution of
  n-inp: An electrochemical impedance study},\ }\href
  {https://doi.org/10.1021/jp0010740} {\bibfield  {journal} {\bibinfo
  {journal} {Journal of Physical Chemistry B}\ }\textbf {\bibinfo {volume}
  {104}},\ \bibinfo {pages} {7725} (\bibinfo {year} {2000})}\BibitemShut
  {NoStop}%
\bibitem [{\citenamefont {Löw}\ \emph {et~al.}(2022)\citenamefont {Löw},
  \citenamefont {Guidat}, \citenamefont {Kim},\ and\ \citenamefont
  {May}}]{Loew_InP_RAS_2022}%
  \BibitemOpen
  \bibfield  {author} {\bibinfo {author} {\bibfnamefont {M.}~\bibnamefont
  {Löw}}, \bibinfo {author} {\bibfnamefont {M.}~\bibnamefont {Guidat}},
  \bibinfo {author} {\bibfnamefont {J.}~\bibnamefont {Kim}},\ and\ \bibinfo
  {author} {\bibfnamefont {M.~M.}\ \bibnamefont {May}},\ }\bibfield  {title}
  {\bibinfo {title} {{The Interfacial Structure of InP(100) in Contact with HCl
  and H$_2$SO$_4$ studied by Reflection Anisotropy Spectroscopy}},\ }\href
  {https://doi.org/10.1039/d2ra05159a} {\bibfield  {journal} {\bibinfo
  {journal} {RSC Advances}\ }\textbf {\bibinfo {volume} {12}},\ \bibinfo
  {pages} {32756} (\bibinfo {year} {2022})}\BibitemShut {NoStop}%
\bibitem [{\citenamefont {Aspnes}(1985)}]{Aspnes1985}%
  \BibitemOpen
  \bibfield  {author} {\bibinfo {author} {\bibfnamefont {D.~E.}\ \bibnamefont
  {Aspnes}},\ }\bibfield  {title} {\bibinfo {title} {{Above-bandgap optical
  anisotropies in cubic semiconductors: A visible--near ultraviolet probe of
  surfaces}},\ }\href {https://doi.org/10.1116/1.582974} {\bibfield  {journal}
  {\bibinfo  {journal} {Journal of Vacuum Science \& Technology B:
  Microelectronics and Nanometer Structures}\ }\textbf {\bibinfo {volume}
  {3}},\ \bibinfo {pages} {1498} (\bibinfo {year} {1985})}\BibitemShut
  {NoStop}%
\bibitem [{\citenamefont {Guidat}\ \emph {et~al.}(2023)\citenamefont {Guidat},
  \citenamefont {Löw}, \citenamefont {Kölbach}, \citenamefont {Kim},\ and\
  \citenamefont {May}}]{Guidat_EC-RAS_review_2023}%
  \BibitemOpen
  \bibfield  {author} {\bibinfo {author} {\bibfnamefont {M.}~\bibnamefont
  {Guidat}}, \bibinfo {author} {\bibfnamefont {M.}~\bibnamefont {Löw}},
  \bibinfo {author} {\bibfnamefont {M.}~\bibnamefont {Kölbach}}, \bibinfo
  {author} {\bibfnamefont {J.}~\bibnamefont {Kim}},\ and\ \bibinfo {author}
  {\bibfnamefont {M.~M.}\ \bibnamefont {May}},\ }\bibfield  {title} {\bibinfo
  {title} {{Experimental and Computational Aspects of Electrochemical
  Reflection Anisotropy Spectroscopy: A Review}},\ }\href
  {https://doi.org/10.1002/celc.202300027} {\bibfield  {journal} {\bibinfo
  {journal} {ChemElectroChem}\ }\textbf {\bibinfo {volume} {10}},\ \bibinfo
  {pages} {e202300027} (\bibinfo {year} {2023})}\BibitemShut {NoStop}%
\bibitem [{\citenamefont {Acosta-Ortiz}\ and\ \citenamefont
  {Lastras-Martínez}(1989)}]{Leo_PRB_1989}%
  \BibitemOpen
  \bibfield  {author} {\bibinfo {author} {\bibfnamefont {S.~E.}\ \bibnamefont
  {Acosta-Ortiz}}\ and\ \bibinfo {author} {\bibfnamefont {A.}~\bibnamefont
  {Lastras-Martínez}},\ }\bibfield  {title} {\bibinfo {title} {{Electro-optic
  effects in the optical anisotropies of (001) GaAs}},\ }\href
  {https://doi.org/10.1103/PhysRevB.40.1426} {\bibfield  {journal} {\bibinfo
  {journal} {Physical Review B}\ }\textbf {\bibinfo {volume} {40}},\ \bibinfo
  {pages} {1426} (\bibinfo {year} {1989})}\BibitemShut {NoStop}%
\bibitem [{\citenamefont
  {Jaegermann}(1996)}]{Jaegermann_modern_asp_elchem_1996}%
  \BibitemOpen
  \bibfield  {author} {\bibinfo {author} {\bibfnamefont {W.}~\bibnamefont
  {Jaegermann}},\ }\bibinfo {title} {{The Semiconductor/Electrolyte Interface:
  A Surface Science Approach}},\ in\ \href
  {http://www.springer.com/chemistry/electrochemistry/book/978-0-306-45450-9}
  {\emph {\bibinfo {booktitle} {{Modern Aspects of Electrochemistry}}}},\
  Vol.~\bibinfo {volume} {30},\ \bibinfo {editor} {edited by\ \bibinfo {editor}
  {\bibfnamefont {R.~E.}\ \bibnamefont {White}}, \bibinfo {editor}
  {\bibfnamefont {B.~E.}\ \bibnamefont {Conway}},\ and\ \bibinfo {editor}
  {\bibfnamefont {J.~O.}\ \bibnamefont {Bockris}}}\ (\bibinfo  {publisher}
  {Plenum Press},\ \bibinfo {address} {New York},\ \bibinfo {year} {1996})\
  Chap.~\bibinfo {chapter} {1}, pp.\ \bibinfo {pages} {1--185}\BibitemShut
  {NoStop}%
\bibitem [{\citenamefont {Tersoff}(1985)}]{Tersoff1985}%
  \BibitemOpen
  \bibfield  {author} {\bibinfo {author} {\bibfnamefont {J.}~\bibnamefont
  {Tersoff}},\ }\bibfield  {title} {\bibinfo {title} {{Schottky barriers and
  semiconductor band structures}},\ }\href
  {https://doi.org/10.1103/PhysRevB.32.6968} {\bibfield  {journal} {\bibinfo
  {journal} {Physical Review B}\ }\textbf {\bibinfo {volume} {32}},\ \bibinfo
  {pages} {6968} (\bibinfo {year} {1985})}\BibitemShut {NoStop}%
\bibitem [{\citenamefont {{May}}\ \emph {et~al.}(2019)\citenamefont {{May}},
  \citenamefont {{Stange}}, \citenamefont {{Weinrich}}, \citenamefont
  {{Hannappel}},\ and\ \citenamefont
  {{Supplie}}}]{May_time-resolved_water_adsorption_2019}%
  \BibitemOpen
  \bibfield  {author} {\bibinfo {author} {\bibfnamefont {M.~M.}\ \bibnamefont
  {{May}}}, \bibinfo {author} {\bibfnamefont {H.}~\bibnamefont {{Stange}}},
  \bibinfo {author} {\bibfnamefont {J.}~\bibnamefont {{Weinrich}}}, \bibinfo
  {author} {\bibfnamefont {T.}~\bibnamefont {{Hannappel}}},\ and\ \bibinfo
  {author} {\bibfnamefont {O.}~\bibnamefont {{Supplie}}},\ }\bibfield  {title}
  {\bibinfo {title} {{The impact of non-ideal surfaces on the solid-water
  interaction: a time-resolved adsorption study}},\ }\href
  {https://doi.org/10.21468/SciPostPhys.6.5.058} {\bibfield  {journal}
  {\bibinfo  {journal} {SciPost Physics}\ }\textbf {\bibinfo {volume} {6}},\
  \bibinfo {pages} {58} (\bibinfo {year} {2019})}\BibitemShut {NoStop}%
\bibitem [{\citenamefont {Vazquez-Miranda}\ \emph {et~al.}(2020)\citenamefont
  {Vazquez-Miranda}, \citenamefont {Solokha}, \citenamefont {Balderas-Navarro},
  \citenamefont {Hingerl},\ and\ \citenamefont
  {Cobet}}]{Vazquez-Miranda_adsorbate_isotherm_RAS_Cu110_HCl_2020}%
  \BibitemOpen
  \bibfield  {author} {\bibinfo {author} {\bibfnamefont {S.}~\bibnamefont
  {Vazquez-Miranda}}, \bibinfo {author} {\bibfnamefont {V.}~\bibnamefont
  {Solokha}}, \bibinfo {author} {\bibfnamefont {R.~E.}\ \bibnamefont
  {Balderas-Navarro}}, \bibinfo {author} {\bibfnamefont {K.}~\bibnamefont
  {Hingerl}},\ and\ \bibinfo {author} {\bibfnamefont {C.}~\bibnamefont
  {Cobet}},\ }\bibfield  {title} {\bibinfo {title} {Adsorbate isotherm analysis
  by reflection anisotropy spectroscopy on copper (110) in hydrochloric acid},\
  }\href {https://doi.org/10.1021/acs.jpcc.9b11326} {\bibfield  {journal}
  {\bibinfo  {journal} {Journal of Physical Chemistry C}\ }\textbf {\bibinfo
  {volume} {124}},\ \bibinfo {pages} {5204} (\bibinfo {year}
  {2020})}\BibitemShut {NoStop}%
\bibitem [{\citenamefont {Euchner}\ \emph {et~al.}()\citenamefont {Euchner},
  \citenamefont {Yadav},\ and\ \citenamefont
  {May}}]{Euchner_phase_diagram_InP_2025}%
  \BibitemOpen
  \bibfield  {author} {\bibinfo {author} {\bibfnamefont {H.}~\bibnamefont
  {Euchner}}, \bibinfo {author} {\bibfnamefont {V.}~\bibnamefont {Yadav}},\
  and\ \bibinfo {author} {\bibfnamefont {M.~M.}\ \bibnamefont {May}},\
  }\bibfield  {title} {\bibinfo {title} {The inp(100) surface phase diagram:
  From the gas-phase to the electrochemical environment},\ }\href
  {https://doi.org/10.1021/acsami.4c20370} {\bibfield  {journal} {\bibinfo
  {journal} {ACS Applied Materials \& Interfaces}\ }\textbf {\bibinfo {volume}
  {17}},\ \bibinfo {pages} {8601}}\BibitemShut {NoStop}%
\bibitem [{\citenamefont {Ponomarev}\ and\ \citenamefont
  {Peter}(1995)}]{Ponomarev_EIS_InP_1995}%
  \BibitemOpen
  \bibfield  {author} {\bibinfo {author} {\bibfnamefont {E.}~\bibnamefont
  {Ponomarev}}\ and\ \bibinfo {author} {\bibfnamefont {L.}~\bibnamefont
  {Peter}},\ }\bibfield  {title} {\bibinfo {title} {A comparison of intensity
  modulated photocurrent spectroscopy and photoelectrochemical impedance
  spectroscopy in a study of photoelectrochemical hydrogen evolution at
  p-inp},\ }\href
  {https://doi.org/http://dx.doi.org/10.1016/0022-0728(95)04148-9} {\bibfield
  {journal} {\bibinfo  {journal} {Journal of Electroanalytical Chemistry}\
  }\textbf {\bibinfo {volume} {397}},\ \bibinfo {pages} {45} (\bibinfo {year}
  {1995})}\BibitemShut {NoStop}%
\bibitem [{\citenamefont {Kyser}\ and\ \citenamefont {Rehn}()}]{Kyser1970}%
  \BibitemOpen
  \bibfield  {author} {\bibinfo {author} {\bibfnamefont {D.}~\bibnamefont
  {Kyser}}\ and\ \bibinfo {author} {\bibfnamefont {V.}~\bibnamefont {Rehn}},\
  }\bibfield  {title} {\bibinfo {title} {Piezoelectric effects in
  electroreflectance},\ }\href {https://doi.org/10.1016/0038-1098(70)90714-3}
  {\bibfield  {journal} {\bibinfo  {journal} {Solid State Communications}\
  }\textbf {\bibinfo {volume} {8}},\ \bibinfo {pages} {1437}}\BibitemShut
  {NoStop}%
\bibitem [{\citenamefont {Baldelli}()}]{Baldelli2005}%
  \BibitemOpen
  \bibfield  {author} {\bibinfo {author} {\bibfnamefont {S.}~\bibnamefont
  {Baldelli}},\ }\bibfield  {title} {\bibinfo {title} {Probing electric fields
  at the ionic liquid--electrode interface using sum frequency generation
  spectroscopy and electrochemistry},\ }\href
  {https://doi.org/10.1021/jp052913r} {\bibfield  {journal} {\bibinfo
  {journal} {Journal of Physical Chemistry B}\ }\textbf {\bibinfo {volume}
  {109}},\ \bibinfo {pages} {13049}}\BibitemShut {NoStop}%
\bibitem [{\citenamefont {Yang}\ \emph {et~al.}(2015)\citenamefont {Yang},
  \citenamefont {Yan}, \citenamefont {Yang}, \citenamefont {Choi},
  \citenamefont {Zhu}, \citenamefont {Luther},\ and\ \citenamefont
  {Beard}}]{Yang_surface_recombination_solution-grown_perovskite_single-crystal_2015}%
  \BibitemOpen
  \bibfield  {author} {\bibinfo {author} {\bibfnamefont {Y.}~\bibnamefont
  {Yang}}, \bibinfo {author} {\bibfnamefont {Y.}~\bibnamefont {Yan}}, \bibinfo
  {author} {\bibfnamefont {M.}~\bibnamefont {Yang}}, \bibinfo {author}
  {\bibfnamefont {S.}~\bibnamefont {Choi}}, \bibinfo {author} {\bibfnamefont
  {K.}~\bibnamefont {Zhu}}, \bibinfo {author} {\bibfnamefont {J.~M.}\
  \bibnamefont {Luther}},\ and\ \bibinfo {author} {\bibfnamefont {M.~C.}\
  \bibnamefont {Beard}},\ }\bibfield  {title} {\bibinfo {title} {{Low surface
  recombination velocity in solution-grown CH$_3$NH$_3$PbBr$_3$ perovskite
  single crystal}},\ }\href {https://doi.org/10.1038/ncomms8961} {\bibfield
  {journal} {\bibinfo  {journal} {Nature Communications}\ }\textbf {\bibinfo
  {volume} {6}},\ \bibinfo {pages} {7961} (\bibinfo {year} {2015})}\BibitemShut
  {NoStop}%
\bibitem [{\citenamefont {Flieg}\ \emph {et~al.}(2025)\citenamefont {Flieg},
  \citenamefont {Guidat},\ and\ \citenamefont {May}}]{Flieg_dataset_2025}%
  \BibitemOpen
  \bibfield  {author} {\bibinfo {author} {\bibfnamefont {M.}~\bibnamefont
  {Flieg}}, \bibinfo {author} {\bibfnamefont {M.}~\bibnamefont {Guidat}},\ and\
  \bibinfo {author} {\bibfnamefont {M.~M.}\ \bibnamefont {May}},\ }\href
  {https://doi.org/10.5281/zenodo.XXXX} {\bibinfo {title} {{Dataset for
  ``Direct observation of potential-dependent surface state formation at an
  electrochemical interface via the optical anisotropy''}}} (\bibinfo {year}
  {2025})\BibitemShut {NoStop}%
\bibitem [{\citenamefont {Schmitt}\ \emph {et~al.}(2023)\citenamefont
  {Schmitt}, \citenamefont {Guidat}, \citenamefont {Nusshör}, \citenamefont
  {Renz}, \citenamefont {Möller}, \citenamefont {Flieg}, \citenamefont
  {Lörch}, \citenamefont {Kölbach},\ and\ \citenamefont
  {May}}]{Schmitt_PEC_Schlenk_cell_2023}%
  \BibitemOpen
  \bibfield  {author} {\bibinfo {author} {\bibfnamefont {E.~A.}\ \bibnamefont
  {Schmitt}}, \bibinfo {author} {\bibfnamefont {M.}~\bibnamefont {Guidat}},
  \bibinfo {author} {\bibfnamefont {M.}~\bibnamefont {Nusshör}}, \bibinfo
  {author} {\bibfnamefont {A.-L.}\ \bibnamefont {Renz}}, \bibinfo {author}
  {\bibfnamefont {K.}~\bibnamefont {Möller}}, \bibinfo {author} {\bibfnamefont
  {M.}~\bibnamefont {Flieg}}, \bibinfo {author} {\bibfnamefont
  {D.}~\bibnamefont {Lörch}}, \bibinfo {author} {\bibfnamefont
  {M.}~\bibnamefont {Kölbach}},\ and\ \bibinfo {author} {\bibfnamefont
  {M.~M.}\ \bibnamefont {May}},\ }\bibfield  {title} {\bibinfo {title}
  {{Photoelectrochemical Schlenk cell funcionalization of multi-junction
  water-splitting photoelectrodes}},\ }\href
  {https://doi.org/10.1016/j.xcrp.2023.101606} {\bibfield  {journal} {\bibinfo
  {journal} {Cell Reports Physical Science}\ }\textbf {\bibinfo {volume} {4}},\
  \bibinfo {pages} {101606} (\bibinfo {year} {2023})}\BibitemShut {NoStop}%
\end{thebibliography}
\end{document}